\documentclass[a4paper,12pt]{article}
\pdfoutput=1
\usepackage{amsfonts,amsthm,amsmath,amssymb,graphicx,hyperref,youngtab}

\newcommand{\be}{\begin{equation}}
\newcommand{\bea}{\begin{eqnarray}}
\newcommand{\ee}{\end{equation}}
\newcommand{\eea}{\end{eqnarray}}

\begin{document}

\makeatletter
\@addtoreset{equation}{section}
\makeatother
\renewcommand{\theequation}{\thesection.\arabic{equation}}

\vspace{1.8truecm}

\vspace{15pt}


{\LARGE{ 
\centerline{\bf From Spinning Primaries to Permutation Orbifolds } 
}}  

\vskip.5cm 

\thispagestyle{empty} 
\centerline{ {\large\bf Robert de Mello Koch$^{a,b,}$\footnote{{\tt robert@neo.phys.wits.ac.za}}, 
Phumudzo Rabambi$^{b,}$\footnote{{\tt 457990@students.wits.ac.za}} }}

\centerline{ {\large\bf and Hendrik J.R. Van Zyl${}^{b,}$\footnote{ {\tt hjrvanzyl@gmail.com}}}}

\vspace{.4cm}
\centerline{{\it ${}^a$ School of Physics and Telecommunication Engineering},}
\centerline{{ \it South China Normal University, Guangzhou 510006, China}}

\vspace{.4cm}
\centerline{{\it ${}^b$ National Institute for Theoretical Physics,}}
\centerline{{\it School of Physics and Mandelstam Institute for Theoretical Physics,}}
\centerline{{\it University of the Witwatersrand, Wits, 2050, } }
\centerline{{\it South Africa } }

\vspace{1truecm}

\thispagestyle{empty}

\centerline{\bf ABSTRACT}

\vskip.2cm 

We carry out a systematic study of primary operators in the conformal field theory of a free Weyl fermion.
Using $SO(4,2)$ characters we develop counting formulas for primaries constructed using a fixed number of
fermion fields.
By specializing to particular classes of primaries, we derive very explicit formulas giving the generating functions 
for the number of primaries in these classes.
We present a duality map between primary operators in the fermion field theory and polynomial functions.
This allows us to construct the primaries that were counted.
Next we show that these classes of primary fields correspond to polynomial functions on certain permutation orbifolds.
These orbifolds have palindromic Hilbert series.

\setcounter{page}{0}
\setcounter{tocdepth}{2}
\newpage
\tableofcontents
\setcounter{footnote}{0}
\linespread{1.1}
\parskip 4pt

{}~
{}~

\section{Introduction}

The remarkable success of the conformal bootstrap\cite{Ferrara:1973yt,Polyakov:1974gs,Mack:1975jr,Rattazzi:2008pe}
 suggests that algebraic structures present in conformal field theory (CFT) can profitably be exploited to extract highly 
nontrivial information about the CFT.
In the papers \cite{Koch:2014nka,Koch:2015pwu} a systematic approach towards manifesting and exploiting 
some of these algebraic structures was outlined.
The key result is that the algebraic structure of CFT defines a two dimensional topological field theory (TFT2)
with $SO(4,2)$ invariance. 
Crossing symmetry is expressed as associativity of the algebra of local CFT operators. 
A basic observation which is at the heart of this result, is that the free four dimensional CFT of a scalar field can be 
formulated as an infinite dimensional associative algebra.
This algebra admits a decomposition into linear representations of $SO(4,2)$, and is equipped with a 
non-degenerate bilinear product. 
A concrete application of these ideas has enabled a systematic study of primaries in bosonic free field theories in 
four dimensions, for scalar, vector and matrix models\cite{deMelloKoch:2017dgi,deMelloKoch:2017caf}.
For closely related ideas see  \cite{Henning:2015daa,Henning:2017fpj}.

We know from the AdS/CFT correspondence\cite{Maldacena:1997re,Gubser:1998bc,Witten:1998qj}
 that strongly coupled CFTs have a dual holographic gravitational description.
The combinatorics of the matrix model Feynman diagrams plays an important role in this holography.
In this setting the TFT2 structure also appears as a powerful organizing structure, explicating algebraic structures
that were not previously appreciated\cite{deMelloKoch:2011uq,Pasukonis:2013ts,Koch:2010zza,Geloun:2013kta}.
Thus, it seems that the TFT2 idea is rich enough to incorporate the algebraic structure emerging both from the
conformal symmetry, and from the color combinatorics.

In this paper we extend the study of \cite{deMelloKoch:2017dgi,deMelloKoch:2017caf} 
by carrying out a systematic study of primaries in free fermion field theories in four dimensions. 
In section \ref{count} we obtain formulae for the counting of primary fields constructed from $n$ copies of a left
handed Weyl fermion, using the characters of representations of $so(4, 2)$. 
For a beautiful discussion of these characters, see \cite{Dolan05}.
The basic quantity that we are interested in is the generating function
\bea
   G_n(s,x,y)=\sum_{\Delta,j_1,j_2}N_{[\Delta,j_1,j_2]}s^\Delta x^{j_1}y^{j_2}\, ,
\eea
which counts the number of conformal multiplets (denoted $N_{[\Delta,j_1,j_2]}$) labeled by the quantum
numbers $\Delta,j_1,j_2$ of their highest weight state.
These quantum numbers are charges of the Cartan of $SO(4,2)$, namely the scaling dimension $\Delta$ and the two spins 
$j_1$ and $j_2$ associated to the $SO(4)=SU(2)\times SU(2)$ subgroup of $SO(4, 2)$.
Although we obtain a concrete expression for $G_n(s,x,y)$, it is not very useful.
By specializing to particular classes of primaries, we can make the counting formulae very explicit.
These special classes of primaries obey extremality conditions stated using relations between the charges 
under the Cartan of $SO(4, 2)$.
The first class of primaries that we consider are the leading twist primaries.
Recall that the twist $\tau$ is given by $\tau = \Delta -(j_1+j_2)$.
As we explain below, the primary operators constructed using $n$ fields that maximize the twist $\tau$
have quantum numbers given by
\bea
    \left[\Delta,j_1,j_2\right]=\left[{n(n+2)\over 2}+q,{n(n+1)\over 4}+{q\over 2},{n(n-1)\over 4}+{q\over 2}
\right]\, .\label{LTQN}
\eea
These quantum numbers are not at all obvious.
To get some insight into the above list, write the scaling dimension as $\Delta ={n(n-1)\over 2}+q+{3\over 2}n$.
The terms $q+{3\over 2}n$ are the expected contribution to the dimension from $q$ derivatives and $n$ fermion fields.
Recall that for scalar fields we'd simply have $\Delta =q+n$ for the leading twist primaries.
Fermi statistics requires that we anti-symmetrize the fermion fields.
Since each field has two components, to get a non-zero answer extra derivatives are needed 
and this leads to the additional contribution of ${n(n-1)\over 2}$.
We denote the generating function counting this class of primaries by $G_n^{\rm max}(s,x,y)$ and we find
\bea
G_n^{\rm max}(s,x,y)=
(s \sqrt{x y})^{n (n-1)\over 2}
(s^{3\over 2}\sqrt{x})^n \prod_{k=2}^n {1\over 1-(s \sqrt{x  y})^k}\, .\label{LTCounttint}
\eea
Following \cite{deMelloKoch:2017dgi,deMelloKoch:2017caf}, we consider a second larger class of primaries, called
the extremal primary operators in \cite{deMelloKoch:2017dgi,deMelloKoch:2017caf}.
This class is the set of operators with maximal $j_1$ spin at given $\Delta$
\bea
    [\Delta,j_1]=[{3n\over 2}+q,{n\over 2}+{q\over 2}]\, .
\eea
We denote the corresponding generating function by $G_n^{\rm ext}(s,x,y)$.
Although we do not have a closed formula for $G_n^{\rm ext}(s,x,y)$ valid for any $n$, we explain how it can be
computed for low values of $n$, by specializing the general counting formula.
As an example we evaluate
\bea
G^{\rm ext}_3(s,x,y) 
&=&{s^{13\over 2}x^{5\over 2}(1+s \sqrt{x} y^{3\over 2})\over 
(1-s^4 x^2)(1-s^2 x y)(1-s^3 x^{3\over 2}y^{3\over 2})}\cr\cr\cr
&=&  s^{13\over 2}x^{5\over 2} + s^{15\over 2}x^3 y^{3\over 2} +s^{17\over 2} x^{7\over 2} y 
+s^{19\over 2} x^4 y^{3\over 2}
+s^{19\over 2} x^4 y^{5\over 2}   + s^{21\over 2} x^{9\over 2} \cr\cr
&& + s^{21\over 2} x^{9\over 2} y^2 + s^{21\over 2} x^{9\over 2} y^3
+\cdots\, .
\eea

After developing these results for the counting of primary operators, we consider the problem of constructing
the primaries that were counted.
The construction of primary fields is mapped to a problem of determining multi-variable polynomials subject to a 
system of algebraic and differential constraints.
Each primary operator corresponds to a specific polynomial.
This relies on a function space realization of the conformal algebra, which is explained in section \ref{FSrealization}.
The special classes of primary operators that we count above have a natural interpretation in this polynomial
construction.
Leading twist primaries correspond to holomorphic polynomials in a single complex variable $z$, while extremal
primaries correspond to holomorphic polynomials in two complex variables, $z$ and $w$. 
We give concrete examples of polynomials obeying the constraints and the associated primary operators.

Finally, in the last section we verify that the Hilbert series for the counting of extremal primaries are palindromic.
The palindromy property of Hilbert series is indicative that the ring being enumerated is Calabi-Yau.
It it interesting that palindromic  Hilbert series also arise for moduli spaces of supersymmetric vacua of gauge theories,
as found in \cite{Gray:2008yu,Hanany:2008sb}.

\section{Counting Primaries}\label{count}

This section considers the problem of enumerating the $SO(4,2)$ irreducible representations appearing among the 
composite fields made out of $n=2,3, \cdots$ copies of a free chiral fermion field.
The chiral fermion is a lowest weight representation with $\Delta={3\over 2}$, $j_1={1\over 2}$ and $j_2=0$.
The fermions are Grassman fields, so there is a sign change when two fields are swapped. 
Consequently, we should be taking the antisymmetric product of the $SO(4,2)$ representations.
We will denote the lowest weight representation corresponding to local operators built by taking derivatives
of the fermion field by $W_+$.
Enumerating the primaries entails decomposing the antisymmetrized tensor product ${\rm Asym}^n (W_+)$
into irreducible representations.
We start by deriving a formula for the character of the antisymmetrized tensor product of $n$ copies of the free Weyl 
fermion representation.
We then explain how to express this character as a sum of characters of irreducible representations, achieving
the required decomposition.
After obtaining a general formula in terms of an infinite product, we specialize to primaries that obey extremality
conditions relating their dimension to their spin. 
For these primaries using results from \cite{BHR2}, we find simple explicit formulas for the counting. 

\subsection{Generalities}

The basic formula we use in this section states
\bea
\det (1+tM)=\sum_{n=0}^\infty  t^n \chi_{(1^n)}(M)\,, \label{basic}
\eea
where $\chi_{(1^n)}(M)$ is the trace over the antisymmetrized product of $n$ copies of $M$.
Below we will use this formula to obtain the character of the antisymmetrized tensor products of $n$ copies of the free 
Weyl fermion representation.
We will reserve the letter $\chi$ for characters.
The character for the free fermion representation is denoted by ${\cal D}_{[{3\over 2},{1\over 2}]+}$ in \cite{Dolan05}. 
From formula (3.44) of  \cite{Dolan05} we know the character of a left handed Weyl fermion is
\bea
   \chi_{W_+}(s,x,y)&=& s^{3\over 2}(\chi_{1\over 2}(x)-s\chi_{1\over 2}(y))P(s,x,y)\cr
                             &=&s^{3\over 2}\sum_{q=0}^\infty s^q\chi_{q+1\over 2}(x)\chi_{q\over 2}(y)\cr
                             &=&{\rm Tr}_{W_+}(M)\,,
\eea
with $M=s^D x^{J_1^3} y^{J_2^3}$  and
\bea
  P(s,x,y)= {1\over (1-s\sqrt{xy})(1-s\sqrt{x\over y})(1-s\sqrt{y\over x})(1-{s\over\sqrt{xy}})}\,.
\eea
Here $J_1^3$ is the third component of the $\vec J_1$ spin. 
It is straightforward to verify that for $M=s^D x^{J_1^3} y^{J_2^3}$ we have
\bea
{\rm det}(1+tM)=\prod_{q=0}^\infty\prod_{a=-{q+1\over 2}}^{q+1\over 2}\prod_{b=-{q\over 2}}^{q\over 2}
(1+ts^{{3\over 2}+q}x^a y^b)\, .
\eea
Applying (\ref{basic}) we find the generating function of the characters of the antisymmetrized tensor products of the
free Weyl fermion representation
\bea
{\cal Z}(t,s,x,y)=\prod_{q=0}^\infty\prod_{a=-{q+1\over 2}}^{q+1\over 2}\prod_{b=-{q\over 2}}^{q\over 2}
(1+ts^{{3\over 2}+q}x^a y^b)
=\sum_{n=0}^\infty  t^n \chi_{(1^n)}(s,x,y)\, .
\eea
By expanding ${\cal Z}(t,s,x,y)$ as a series in $t$ we can easily read off  $\chi_{(1^n)}(s,x,y)$ as the coefficient of $t^n$.
To be clear, $\chi_{(1^n)}(s,x,y)$ is the character of $M$ in the representation given by the
antisymmetrized tensor product ${\rm Asym}^n (W_+)$.
The next step is to decompose this into a sum of $SO(4,2)$ characters, for irreps of dimension $\Delta$ and spins
$j_1,j_2$
\bea
\chi_{(1^n)}(s,x,y)=\sum_{[\Delta,j_1,j_2]}N_{[\Delta,j_1,j_2]}\chi_{[\Delta,j_1,j_2]}(s,x,y)\,.
\eea
The coefficients $N_{[\Delta,j_1,j_2]}$ count how many times the irreducible representation with lowest weight labeled
by $[\Delta,j_1,j_2]$ appears in ${\rm Asym}^n (W_+)$.
Hence, $N_{[\Delta,j_1,j_2]}$ are non-negative integers.
The case that $n=2$ is complicated by the fact that some of the irreducible representations appearing in the above 
decomposition are short.
We will consider $n=2$ separately in detail below. 
For $n\ge 3$ the character for the irreducible representation with lowest weight $[\Delta,j_1,j_2]$ is given by\cite{Dolan05}
\bea
   \chi_{[\Delta,j_1,j_2]}(s,x,y)=
{s^\Delta\chi_{j_1}(x)\chi_{j_2}(y)\over (1-s\sqrt{xy})(1-s\sqrt{x\over y})(1-s\sqrt{y\over x})(1-{s\over\sqrt{xy}})}\,.
\eea
It is useful to define 
\bea\label{eq:partition}
Z_n (s,x,y)&\equiv& \sum_{\Delta,j_1,j_2}N_{[\Delta,j_1,j_2]}s^\Delta\chi_{j_1}(x)\chi_{j_2}(y)\,,
\eea
so that
\bea 
Z_n ( s, x,  y ) =
(1-s\sqrt{xy})(1-s\sqrt{x\over y})(1-s\sqrt{y\over x})(1-{s\over\sqrt{xy}}) ~ \chi_{(1^n)}(s,x,y) \, .
\label{forZ}
\eea
The right hand side of (\ref{eq:partition}) is a sum of (products of) $SU(2)$ characters.
Following \cite{spradlin}, it can be simplified by using the orthogonality of $SU(2)$ characters.
Towards this end, we introduce the generating function
\bea
  G_n (s,x,y)&\equiv&\sum_{\Delta,j_1,j_2}N_{[\Delta,j_1,j_2]}s^\Delta x^{j_1}y^{j_2}\cr
&=& \left[ (1-{1\over x})(1-{1\over y})Z_n (s,x,y)\right]_{\ge}\,\, .
\eea
The subscript $\ge$ is an instruction to keep only non negative powers of $x$ and $y$.

It is easy to check that this agrees with standard character computations. 
For example, the expansion
\bea
G_3(s,x,y)&=&s^{11\over 2}x\sqrt{y}+s^{13\over 2}x^{5\over 2}+s^{13\over 2} x^{3\over 2} y
+s^{15\over 2}y^{3\over 2}+s^{15\over 2}x^3 y^{3\over 2}+s^{15\over 2} x^2 y^{3\over 2}
+s^{17\over 2} x^{7\over 2} y\cr
&+&s^{17\over 2} x^{3\over 2} y^2+s^{17\over 2} x^{5\over 2} y^2
+s^{19\over 2} x^4 y^{3\over 2}+s^{19\over 2} x y^{5\over 2}+2 s^{19\over 2} x^3 y^{5\over 2}
+s^{19\over 2} x^4 y^{5\over 2}+....\,,\cr
&&\label{nisthreeexamplefrt}
\eea
can be reproduced using characters, as we will now demonstrate.
The relevant Schur polynomial for this case is calculated as follows
\bea
\chi_{(1^{3})}(s,x,y) 
= {1 \over 6} \left[ (\chi_{L}(s,x,y))^{3} - 3\chi_{L}(s^{2},x^{2},y^{2})\chi_{L}(s,x,y) 
+ 2\chi_{L}(s^{3},x^{3},y^{3}) \right]\, .\label{nisthreeexample}
\eea
Using Mathematica, we find the following terms
\bea
\chi_{(1^{3})}(s,x,y) &=& \chi_{[{11 \over 2},1,{1 \over 2}]}(s,x,y)+\chi_{[{13 \over 2},{5 \over 2},0]}(s,x,y)
+\chi_{[{13 \over 2},{3 \over 2},1]}(s,x,y)\cr\cr
&+&\chi_{[{15 \over 2},0,{3 \over 2}]}(s,x,y)+\chi_{[{15 \over 2},2,{3 \over 2}]}(s,x,y)
+\chi_{[{15 \over 2},3,{3 \over 2}]}(s,x,y) \cr\cr
&+& \chi_{[{17 \over 2},{7 \over 2},1]}(s,x,y)+\chi_{[{17 \over 2},{3 \over 2},2]}(s,x,y)
+\chi_{[{17 \over 2},{5 \over 2},2]}(s,x,y)\cr\cr
&+&\chi_{[{19 \over 2},4,{3 \over 2}]}(s,x,y)+\chi_{[{19 \over 2},1,{5 \over 2}]}(s,x,y)
+2\chi_{[{19 \over 2},3,{5 \over 2}]}(s,x,y)+\chi_{[{19 \over 2},4,{5 \over 2}]}(s,x,y)\cr\cr
&+&\chi_{[{21 \over 2},{9 \over 2},0]}(s,x,y)+\chi_{[{21 \over 2},{9 \over 2},2]}(s,x,y)
+\chi_{[{21 \over 2},{3 \over 2},3]}(s,x,y)+\chi_{[{21 \over 2},{5 \over 2},3]}(s,x,y)\cr\cr
&+&\chi_{[{21 \over 2},{7 \over 2},3]}(s,x,y)+\chi_{[{21 \over 2},{9 \over 2},3]}(s,x,y)+....\,,
\eea
in complete agreement with (\ref{nisthreeexamplefrt}).

To end this subsection, we will now discuss the case that $n=2$.
For this case we must account for the fact that representations that include null states appear in the decomposition.
A lowest weight multiplet $[\Delta,j_1,j_2]$ will be short  if \cite{Minwalla:1997ka} $\Delta=f(j_1)+f(j_2)$ with 
$f(j)=0$ if $j=0$ or $f(j)=j+1$ if $j>0$.
This does not cover the case of the scalar field ($j_1=j_2=0$), which is short for $\Delta =1$.
For $n=2$ the decomposition includes a primary with $\Delta =3$ and $j_1=j_2=0$ which is not short, as well as
primaries with $\Delta = 2j$ $j_1=(2j-1)/2$ and $j_2=(2j-3)/2$ which are short representations and hence have null states.
For the correct counting, these null states (and their descendants) must be removed.
These short representations arise because their primary operators are conserved higher spin currents 
\bea
   \partial_\mu \,  J^{\mu\mu_2\cdots\mu_{2j-2}}_{\alpha\dot\beta}=0\, .
\eea
The subtraction of null states is achieved by removing the $\Delta=3$ primary that does not need to be subtracted,
dividing by $1-s/\sqrt{xy}$ which removes the null descendents  and then putting the original primary back in.
The final result
\bea
  G_2 (s,x,y)&=&\left[ 
(1-{1\over x})(1-{1\over y})\left( Z_2 (s,x,y) -s^3\right)
{1\over 1-{s\over\sqrt{xy}}}
\right]_{\ge}+s^3\cr
&=&\sum_{j=0}^\infty  s^{3+2j}x^{{3\over 2}+j} y^{{1\over 2}+j}\,,
\eea
agrees with \cite{Flato:1978qz}.

\subsection{Leading Twist Primaries}

By restricting to well defined classes of primaries, we can significantly simplify the counting formulas of the previous
section.
The biggest simplification comes from focusing on the leading twist primaries, which have quantum numbers
$[\Delta,j_1,j_2]=[{n(n+2)\over 2}+q,{n(n+1)\over 4}+{q\over 2},{n(n-1)\over 4}+{q\over 2}]$.
These quantum numbers are not obvious but will be evident in the final answer of this section.
In the introduction we motivated these quantum numbers, in the discussion appearing after equation (\ref{LTQN}).
Each such leading twist primary operator comes in a complete spin multiplet of 
$({n(n+1)\over 2}+q+1)({n(n-1)\over 2}+q+1)$ operators.
Choosing the operator with highest spin corresponds to studying primaries constructed using a single component  
$P_z$ of the momentum four vector operator.
To count the leading twist primaries we will count this highest spin operator in each multiplet.
The corresponding generating function is $G_n^{\rm max} (s,x,y)$.
This generating function is obtained after a simple modification of the results of the previous section.
First, we replace $\chi_{(1^n)}(s,x,y)$ with a new function
$\chi^{\rm max}_n (s,x,y)$, by keeping only the highest spin state from each multiplet in the product
\bea
\prod_{q=0}^\infty
(1+ts^{{3\over 2}+q}x^{{q\over 2}+{1\over 2}} y^{q\over 2})
=\sum_{n=0}^\infty t^n \chi^{\rm max}_n (s,x,y)\, .
\eea
The leading twist primaries are constructed using the single component of the momentum that raises  
left and right spin maximally. Consequently in (\ref{forZ}) we replace
\bea
(1-s\sqrt{xy})(1-s\sqrt{x\over y})(1-s\sqrt{y\over x})(1-{s\over\sqrt{xy}})\to (1-s\sqrt{xy})\, .
\eea
Finally, for each spin multiplet we keep only 1 state so there is no longer any need to replace the multiplet of spin
states by a single state when we count. 
The final result is
\bea
G_n^{\rm max} (s,x,y) &\equiv& \sum_{\Delta,j_1,j_2}N^{\rm max}_{[\Delta,j_1,j_2]}s^\Delta x^{j_1}y^{j_2}\cr
&=& (1-s\sqrt{xy})\chi^{\rm max}_n (s,x,y)\,,
\eea
where $N^{\rm max}_{[\Delta,j_1,j_2]}$ is the number of leading twist primaries of dimension $\Delta$ and
spin $(j_1,j_2)$.
For the leading twist primaries, once $n$ and the dimension of the operator is specified, the spin of the primary is fixed.
Consequently, we need not track the $x$ and $y$ dependence, although we choose to keep this dependence explicit.
This leads to the formula
\bea
F(t,s,x,y)&\equiv&{1\over (1-s\sqrt{xy})}\sum_{n=0}^\infty t^n G_n^{\rm max} (s,x,y)\cr
&=& \prod_{q=0}^\infty (1+ts^{{3\over 2}+q}x^{{q\over 2}+{1\over 2}} y^{q\over 2})
\, .
\eea
We can obtain explicit expressions for $G_n^{\rm max} (s,x,y)$ by developing $F(t,s,x,y)$ in a Taylor series.
Define
\bea
   f_k (t,s,x,y)={\partial^k\over\partial t^k} \log F(t,s,x,y)\, .
\eea
A straight forward computation gives
\bea
f_k (t,s,x,y) =\sum_{q=0}^{\infty} 
{(-1)^{k+1} (k-1)! s^{{3k\over 2}+kq}x^{{kq\over 2}+{k\over 2}}y^{kq\over 2}
\over (1+t s^{{3\over 2}+q}x^{{q\over 2}+{1\over 2}} y^{q\over 2})^k}\,,
\eea
so that we have
\bea
f_k(0,s,x,y)=(k-1)!(-1)^{k-1}
{s^{3k\over 2}x^{k\over 2} \over 1-s^{k}x^{k\over 2} y^{k\over 2}}\, .
\eea
Explicit expressions for $G_n^{\rm max}$ are now easily obtained.
For example
\bea
G_3^{\rm max}(s,x,y)&=&{1\over 3!}(1-s\sqrt{xy}){\partial^3 F\over\partial t^3}\Big|_{t=0}\cr
&=&{1\over 3!}(1-s\sqrt{xy})(f_3+3f_1 f_2+f_1^3)\cr
&=&{s^{15\over 2}x^{3}y^{3\over 2}\over (1-s^2 xy)(1-s^3x^{3\over 2}y^{3\over 2})}\, .
\eea
Similarly
\bea
G_4^{\rm max}(s,x,y)={s^{12}x^5 y^3\over (1-s^2 xy)(1-s^3x^{3\over 2}y^{3\over 2})(1-s^4 x^2 y^2)}\, .
\eea
It is possible to obtain a general closed formula for $G_n^{\rm max}(s)$.
To make the argument as transparent as possible, set $x=1=y$.
Evaluate the derivative
\bea
   {\partial^n F\over\partial t^n}=
\sum_{n_1,\cdots, n_q}\sum_{k_1,\cdots, k_q}
{(n_1k_1+\cdots+n_q k_q)!\over n_1!\cdots n_q!(k_1!)^{n_1}\cdots (k_q!)^{n_q}}
f_{k_1}^{n_1}\cdots f^{n_q}_{k_q}\delta_{n,n_1k_1+\cdots n_q k_q}\,\, F\,,
\eea
and use the formulas for the $f_k$'s to find
\bea
   {\partial^n F\over\partial t^n}\Big|_{t=0}
&=&\sum_{n_1,\cdots, n_q}\sum_{k_1,\cdots, k_q}
{(-1)^{n-\sum_i n_i}n! s^{3n\over 2}\over n_1!\cdots n_q!\, k_1^{n_1}\cdots k_q^{n_q}}
\big({s^{3k_1\over 2} \over 1-s^{k_1}}\big)^{n_1}\cdots 
\big({s^{3k_q\over 2}\over 1-s^{k_q}}\big)^{n_q}
\delta_{n,n_1k_1+\cdots n_q k_q}\,.\cr
&&\label{genformI}
\eea
The sum appearing above can be interpreted as a sum over conjugacy classes of $S_n$.
Recall that a conjugacy class of $S_n$ collects all permutations with $n_q$ $k_q$-cycles, that is, all permutations
with the same cycle structure.
This identification of the sum is a consequence of the fact that the coefficient
\bea
{n!\over n_1!\cdots n_q!\, k_1^{n_1}\cdots k_q^{n_q}}
\eea
is the order of the conjugacy class. 
Each conjugacy class is weighted by the factor $(-1)^{n-\sum_i n_i}$ which is the signature of the permutation
with $n_q$ $k_q$-cycles.
There is a factor of ${s^{3k\over 2}\over 1-s^{k}}$ for each $k$-cycle in the permutation.
The lowest weight discrete series irreducible representation of $SL(2)$, built on a ground state with 
dimension ${3\over 2}$ has character
\bea
   \chi_1(s)={\rm Tr}_{V_1}(s^{L_0})={s^{3\over 2}\over 1-s}\, .
\eea
Denote this irreducible representation by $W_1$.
It then follows that ($P_{[1^n]}$ projects onto the antisymmetric irrep i.e. a single column of $n$ boxes)
\bea
  {1\over n!} {\partial^n F\over\partial t^n}\Big|_{t=0}={\rm Tr}_{W_1}(P_{[1^n]} s^{L_0})=
{s^{{n\over 2}(n+2)}\over (1-s)(1-s^2)(1-s^3)\cdots (1-s^n)}\, .\label{LTCount}
\eea
In writing the last equality above, we have used equation (49) of \cite{BHR2} which studies the $SL(2)$ sector primaries 
using the language of oscillators. 
Our final result, for general $x$ and $y$, is
\bea
G_n^{\rm max}(s,x,y)=
(s \sqrt{x y})^{n (n-1)\over 2}
(s^{3\over 2}\sqrt{x})^n \prod_{k=2}^n {1\over 1-(s \sqrt{x  y})^k}\, .\label{LTCountt}
\eea

\subsection{Extremal Primaries}

In this section we will consider the class of primaries with charges
\bea 
 \Delta = {3n\over 2} + q ~~ ; ~~   J_1^3  = {n\over 2}+{q\over 2}\, .
\eea
This class of primaries generalizes the higher twist primaries because the charge $J_2^3$, which is part of $SU(2)_R$, 
is not constrained. 
The extremal primaries fill out complete multiplets of $SU(2)_R$ and are constructed using two components of the 
momentum four vector operator which are complex linear combinations of the (hermitian) $P_\mu$.
The specific complex linear combinations are determined by the requirement that $J_1^3$ is maximal.
Following the treatment of the last section, we introduce a generating function $G^{\rm ext}_n(s,x,y)$, given by
\begin{eqnarray}
G^{\rm ext}_n(s,x,y)=\left[\left(1-{1\over y}\right)Z^{\rm ext}_n(s,x,y)\right]_{\ge} \label{zwcount}
\end{eqnarray}
where $Z^{\rm ext}_n(s,x,y)$ is defined by
\bea
Z^{\rm ext}_n(s,x,y)=(1-s\sqrt{xy})(1-s\sqrt{x/y})\chi^{\rm ext}_n(s,x,y)\label{forzzw}
\eea
with
\bea
F^{(2)} (t,s,x,y) &\equiv&\sum_{n=0}^{\infty}t^n \chi^{\rm ext}_{n} (s,x,y)\cr
&=& \prod_{q=0}^\infty\prod_{b=-{q\over 2}}^{q\over 2} (1+ts^{{3\over 2}+q}x^{q+1\over 2} y^b)\, .
\label{zwprod}
\end{eqnarray}
It is again possible to derive closed expressions for the generating function $G^{\rm ext}_n(s,x,y)$. 
Introduce the functions
\bea
f_k(t,s,x,y)&\equiv&{\partial^{k-1}\over\partial t^{k-1}}\log F^{(2)}\cr
&=&(-1)^{k-1}(k-1)!\sum_{q=0}^\infty\sum_{m=-{q\over 2}}^{q\over 2} 
{s^{kq+{3k\over 2}}x^{(q+1)k\over 2}y^{km}\over 
(1+t s^{q+{3\over 2}}x^{q+1\over 2}y^{m})^k}\, .
\eea
It is simple to establish that
\bea
f_k(0,s,x,y) = (-1)^{k-1} (k-1)! 
{s^{3k\over 2}x^{k\over 2}\over (1-s^k x^{k\over 2}y^{k\over 2})(1-s^k x^{k\over 2} y^{-{k\over 2}})}\, .
\eea
Exactly as above we have
\bea
   {\partial^n F^{(2)}\over\partial t^n}\Big|_{t=0}=
\sum_{n_1,\cdots, n_q}\sum_{k_1,\cdots, k_q}
{(n_1k_1+\cdots+n_q k_q)!\over n_1!\cdots n_q!(k_1!)^{n_1}\cdots (k_q!)^{n_q}}
f_{k_1}^{n_1}\cdots f^{n_q}_{k_q}\delta_{n,n_1k_1+\cdots n_q k_q}\, .
\eea
Inserting the formulas for the $f_k$'s, expressions for the $Z^{\rm ext}_n(s,x,y)$ now follow from (\ref{forzzw}).
To extract spin multiplets, we need to compute
\bea
G^{\rm ext}_n(z,w)=
\left[ Z_n (s,x,y) \left(1-{1\over y}\right)\right]_{\ge}
={1\over 2\pi i}\oint_C dz
{\left(1-{1\over z^2}\right)Z_n (s,x,z^2)\over z-\sqrt{y}}\, .
\eea
As an example, the generating functions counting the extremal primaries constructed from 3 fields are given by
\bea
Z_3^{\rm ext}(s,x,y)=s^{13\over 2}x^{5\over 2}y^{-{3\over 2}}
{(y^{3\over 2}+s^2 x y^{3\over 2}+s\sqrt{x}(1+y)(1+y^2))\over
(1-s^2 xy)(1-s^3 x^{3\over 2} y^{3\over 2})(1-{s^2 x\over y})(1-{s^3 x^{3\over 2}\over y^{3\over 2}})}
\label{Z3}
\eea
\bea
G^{\rm ext}_3(s,x,y) 
&=&{s^{13\over 2}x^{5\over 2}(1+s \sqrt{x} y^{3\over 2})\over 
(1-s^4 x^2)(1-s^2 x y)(1-s^3 x^{3\over 2}y^{3\over 2})}\cr\cr
&=&  s^{13\over 2}x^{5\over 2} + s^{15\over 2}x^3 y^{3\over 2} +s^{17\over 2} x^{7\over 2} y 
+s^{19\over 2} x^4 y^{3\over 2}
+s^{19\over 2} x^4 y^{5\over 2}   + s^{21\over 2} x^{9\over 2} \cr\cr
&& + s^{21\over 2} x^{9\over 2} y^2 + s^{21\over 2} x^{9\over 2} y^3
+\cdots\, .
\eea

\section{Construction}\label{FSrealization}

In this section we will explain how the counting of the previous section can be used to derive concrete formulas
for the construction of the primary operators in the free fermion CFT.
For the leading twist counting this is manifest.
For the counting of extremal primaries, we will argue that our formulas can naturally be phrased as counting the
multiplicities of symmetric group representations.
The quantities being counted are then easily constructed using projectors onto these representations.
In this analysis, a polynomial representation of $SO(4,2)$ will play an important role.
This representation is described in the next subsection, after which we describe the construction of leading
twist primaries and then extremal primaries.

\subsection{Polynomial rep}

We use the following representation of $SO(4,2)$
\bea
  K_\mu = {\partial\over\partial x^{\mu}}\, ,
\eea
\bea
  D =\Big(x\cdot{\partial\over\partial x}-{3\over 2} \Big)\, ,
\eea
\bea
M_{\mu\nu}= x_\mu{\partial\over\partial x^{\nu}}-x_\nu{\partial\over\partial x^{\mu}}+{\cal M}_{\mu\nu}\, ,
\eea
\bea
P_\mu =(x^{2}{\partial\over\partial x^{\mu}} - 2x_\mu x\cdot {\partial\over\partial x}+3 x_\mu - 2x^{\nu}
{\cal M}_{\mu\nu})\, .\label{forP}
\eea
In the formula above we should replace ${\cal M}_{\mu\nu}$ by the relevant matrix representing the spin part
of the conformal group. 
In Minkowski spacetime we have (the two possibilities correspond to taking either a left handed $({1\over 2},0)$ or a 
right handed $(0,{1\over 2})$ spinor)
\bea
 {\cal  M}^{\mu\nu}= \sigma^{\mu\nu}\,,\qquad {\rm or}\qquad  \bar\sigma^{\mu\nu}\, ,
\eea
where
\bea
  ( \sigma^{\mu\nu})_\alpha{}^\beta
     ={1\over 4}\left(\sigma^\mu\bar\sigma^\nu -\sigma^\nu\bar\sigma^\mu\right)_\alpha{}^\beta\,,
\eea
\bea
  (\bar{\sigma}^{\mu\nu})^{\dot\alpha}{}_{\dot\beta}
  ={1\over 4}\left(\bar\sigma^\mu\sigma^\nu -\bar\sigma^\nu \sigma^\mu\right)^{\dot\alpha}{}_{\dot\beta}\,,
\eea
and
\bea
 \sigma^\mu{}_{\alpha\dot\beta}=({\bf 1},\vec{\sigma})\,,\qquad\qquad 
\bar\sigma^{\mu\dot\beta\alpha}=({\bf 1},-\vec{\sigma})\, .
\eea
In Euclidean space we have
\bea
{\cal  M}^{\mu\nu}= \sigma^{\mu\nu}\equiv 
{1\over 4}\left(\sigma^\mu\bar\sigma^\nu -\sigma^\nu\bar\sigma^\mu\right)\, ,
\eea
or
\bea
{\cal  M}^{\mu\nu}=\bar{\sigma}^{\mu\nu}
\equiv {1\over 4}\left(\bar\sigma^\mu\sigma^\nu -\bar\sigma^\nu \sigma^\mu\right)\, ,
\eea
where now
\bea
 \sigma^\mu=(-i\vec{\sigma},{\bf 1})\,, \qquad\qquad \bar\sigma^{\mu}=(i\vec{\sigma},{\bf 1})\, .
\eea
The generators in Minkowski space close the algebra
\bea
&&[M_{\rho\sigma},M_{\phi\theta}]=\eta_{\theta\sigma}M_{\phi\rho}+\eta_{\phi\rho}M_{\theta\sigma}
-\eta_{\theta\rho}M_{\phi\sigma}-\eta_{\phi\sigma}M_{\theta\rho}\,,\cr\cr
&&[P_\mu,P_\nu]=0=[K_\mu,K_\nu]\,,\qquad
[P_\beta, K_\alpha]=2\eta_{\alpha\beta}D-2M_{\alpha\beta}\,,\cr\cr
&&[M_{\beta\rho},K_\alpha]=\eta_{\alpha\rho}K_\beta-\eta_{\alpha\beta}K_\rho\,,\qquad
[M_{\beta\rho},P_\alpha]=\eta_{\alpha\rho}P_\beta-\eta_{\alpha\beta}P_\rho\,,\cr\cr
&&[D,P_\mu]=P_\mu\,,\qquad [D,K_\mu]=-K_\mu\,,\qquad [D,M_{\mu\nu}]=0\,.
\eea
The Euclidean generators obey the same algebra with $\eta_{\mu\nu}$ replaced with $\delta_{\mu\nu}$.

States in this representation correspond to polynomials in the spacetime coordinates $x^\mu$
times a spinor $\zeta_\alpha$, which is independent of $x^\mu$ and transforms in the $({1\over 2},0)$ 
if we study the theory of a left handed fermion, or in the $(0,{1\over 2})$ if we study a right handed fermion.
The 2$\times$2 matrix ${\cal M}_{\mu\nu}$ acts on this spinor.
Further, $\zeta_\alpha$ is Grassman valued to account for the fact that the fermions are anticommuting fields.
Concretely, each operator corresponds to a state (by the state operator correspondence) and each state corresponds to 
a polynomial times the spinor (thanks to the representation we have just described)
\bea
  x^{\mu_1}\cdots x^{\mu_k}\zeta_\alpha\, .
\eea
To deal with operators constructed from a product of $n$ copies of the basic fermion field, we consider a ``multiparticle
system''. 
When we move to the multiparticle system, we have polynomials  on the $n$ particle coordinates $x_\mu^I$, times the
$n$ particle spinor, obtained by taking the tensor product of $n$ copies of $\zeta_\alpha$
\bea
(\zeta\otimes\zeta\otimes\cdots\otimes\zeta)_{\alpha_1\alpha_2\cdots\alpha_n}\, .
\eea
To write the generator of the conformal group, for the multiparticle system, we need the matrices
\bea
   {\cal M}_{\mu\nu}^{(I)}={\bf 1}\otimes\cdots\otimes
                                          {\bf 1}\otimes {\cal M}_{\mu\nu}\otimes{\bf 1}\otimes\cdots\otimes{\bf 1}
\eea
where the matrix ${\cal M}_{\mu\nu}$ on the right hand side is the 2$\times$2 matrix we introduced above and
it appears as the $I$th factor.
In total  ${\cal M}_{\mu\nu}^{(I)}$ has $n$ factors.
The $n$-particle representation of $SO(4,2)$ includes
\bea
  K_\mu = \sum_{I=1}^n {\partial\over\partial x^I_{\mu}}\,,
\eea
and
\bea
P_\mu =\sum_{I=1}^n ((x^{I\rho}x^I_\rho {\partial\over\partial x^I_{\mu}} - 2x^I_\mu x^I\cdot 
{\partial\over\partial x^I}+ 3 x^I_\mu - 2x^{I\, \nu} {\cal M}^{(I)}_{\mu\nu})\, .
\eea

The representations introduced above all have null states.
This is to be expected, since the dimension of the free fermion field saturates the unitarity bound.
For the $({1\over 2},0)$ field in Minkowski spacetime for example, the null state is exhibited by verifying that
\bea
\bar\sigma^\mu P_\mu \zeta = 0\label{nsc}
\eea
for any choice of $\zeta$.

Let us now spell out the conditions that the polynomial $P_{\cal O}$ corresponding to an operator ${\cal O}$ must obey if the
operator ${\cal O}$ is a primary operator.
The general polynomial $P_{\cal O}$ will have spinor indices (it is constructed from a tensor product of copies of $\zeta$)
as well as four vector indices inherited from the spacetime coordinates.
There are three conditions that must be imposed: Primaries are annihilated by the special conformal generator $K_\mu$
\bea
[ K_\mu ,{\cal O}]=0\, .
\eea
This implies that the corresponding polynomial is translation invariant
\bea
\sum_{I=1}^n{\partial\over\partial x_\mu^I} P_{\cal O} =0\,.
\eea
Secondly, the equation of motion must be obeyed by each fermionic field.
Finally, we require that the polynomials are in the antisymmetric representation of $S_n$. 
Since the $\zeta$s are Grassman variables, we must impose this condition if we are to get a non-zero primary
upon translating back to the language of the fermion field theory.

We do not know how to obtain the complete set of polynomial solutions to the above constraints, 
corresponding to determing the complete set of primaries. 
We can however find a class of solutions and these correspond precisely to the leading twist and
extremal primaries that we counted above.
The fact that these polynomials are to be identified with the leading twist and extremal primaries will be evident
in the detailed match between the counting of these solutions (performed in the following subsections) and the counting 
of the leading twist and extremal primaries.
We will now explain how to find a large class of polynomials that solve the equation of motion constraint, leaving 
the discussion of the remaining two constraints for the subsections which follows.
In the remainder of this subsection, we will work in Euclidean space.
We use $x_4=ix_0$ for the Euclidean time coordinate.

Our first observation is simply that any polynomial in the momenta $P(P_\mu)$, acting on the spinor $\zeta$, solves the
equation of motion constraint. Indeed, since the different components of momentum commute, we know that
\bea
  \bar\sigma^\mu P_\mu \, P(P_\alpha)\zeta = P(P_\alpha) \, \bar\sigma^\mu P_\mu \zeta =0\,,
\eea
with the last equality following from (\ref{nsc}).
Introduce the complex variables
\bea
z= x_2+ix_1,\qquad w=x_3+ix_4,
\eea
and momenta
\bea
P_z= P_2+iP_1, \qquad P_w=P_3+iP_4\, .
\eea
Our second observation is that if we specialize to a $\zeta$ with maximal $J_1^3$ eigenvalue, then any 
polynomial holomorphic in $z$ and $w$ can be translated into a polynomial in $P_z$ and $P_w$.
It is easy to see from a few examples, that $(P_z)^k\zeta \propto z^k\zeta$.
When performing this computation use the identity
\bea
(P_2+iP_1)\zeta =P_z\zeta = z \zeta\,,
\eea
which holds for the spinor with maximal $J_1^3$ eigenvalue.
For our choices above, this spinor is given by
\bea
\zeta =\left[\begin{matrix} 0 \cr 1 \end{matrix}\right]\, .
\eea
Define the number $a_k$ by the relation
\bea
(P_z)^k\zeta = a_k z^k\zeta\, .
\eea
Then
\bea
(P_z)^{k+1} \zeta &=& P_z a_k z^k \zeta\cr
                             &=&-2(k+1)a_k z^{k+1}\zeta\cr
                              &=&a_{k+1}z^{k+1}\zeta\, .
\eea
Thus, we have $a_{k+1}=-2(k+1)a_k$.
This recursion together with the intial value $a_1=-2$, implies that
\bea
   a_k = (-2)^k  k!
\eea
Thus we obtain the following translation between polynomials and momenta
\bea
  (P_z)^k\zeta = (-1)^k 2^k k! \, z^k\zeta\, .
\eea
When peforming this computation note that the first term in (\ref{forP}) does not contribute because the complex
combination we consider assembles the derivative $\partial_{\bar z}$ from this first term.
The last two terms give $-2z$ for the spinor $\zeta$ we are using.
Using a very similar argument, we find
\bea
  (P_z)^k (P_w)^l \zeta = (-2)^{k+l} (k+l)! \, z^k\, w^l\, \zeta
\eea
We can now argue that any polynomial in $z$ and $w$ multiplying the spinor $\zeta$ with maximal $J_1^3$ eigenvalue, 
obeys the equation of motion constraint.
It is enough to argue for a single monomial, since any polynomial is a sum of monomials.
We argue as follows
\bea
\bar\sigma^\mu P_\mu (z^k w^l\zeta) &=&
{1\over (-2)^{k+l}(k+l)!}\bar\sigma^\mu P_\mu (P_z^k P_w^l\zeta)\cr
&=&{1\over (-2)^{k+l}(k+l)!}P_z^k P_w^l(\bar\sigma^\mu P_\mu \zeta)\cr
&=&0
\eea
which demonstrates the claim.

\subsection{Leading Twist}

Using a counting argument, we will confirm that the leading twist primaries are given by 
polynomials in a single complex variable $z^I$, $I=1,2,...,n$.
Any such polynomial obeys the equation of motion constraint.
To solve the translation invariance condition, we work with the hook variables $Z^a$, $a=1,2,...,n-1$ defined by
\bea
Z^a = { 1 \over \sqrt {  a ( a +1 )}  } 
 (z^{(1)}+z^{(2)} + \cdots +z^{(a)} - a z^{(a+1)} )   \, .
\eea
These variables fill out the hook representation of $S_n$, which is labeled by a Young diagram whose first row
has $n-1$ boxes and second row has 1 box.
We denote the corresponding vector space by $V_H$, with the subscript $H$ for ``hook''.
Our problem is now reduced to constructing antisymmetric polynomials from the hook variables.
By construction, it is clear that the degree $k$ polynomials belong to a subspace of $V_H^{\otimes k}$ of $S_n$.
We can characterize the antisymmetric subspace, that we want to extract, using representation theory. 
Towards this end, consider the following decomposition in terms of $S_n \times S_k$ irreps
\bea
V_H^{\otimes k}=\bigoplus_{\Lambda_1\vdash n,\,\, \Lambda_2\vdash k}
V^{(S_n)}_{\Lambda_1} \otimes V^{(S_k)}_{\Lambda_2}\otimes 
V^{Com(S_n\times S_k)}_{\Lambda_1,\Lambda_2}\, .
\eea
In the above expression, $Com(S_n \times S_k)$ is the algebra of linear operators on $V_H^{\otimes k}$ that commute
with $S_n\times S_k$, $V^{(S_n)}_{\Lambda_1}$ carries the irreducible representation $\Lambda_1$ of $S_n$,
$V^{(S_k)}_{\Lambda_2}$ carries the irreducible representation $\Lambda_2$ of $S_k$ and 
$V^{Com(S_n\times S_k)}_{\Lambda_1,\Lambda_2}$ carries the representation $(\Lambda_1,\Lambda_2)$ of 
$Com(S_n\times S_k)$.
This decomposition has been studied in detail in \cite{BHR2}.
The $Z$ variables are commuting so that we need to consider the case that $\Lambda_2=[k]$, the symmetric representation
given by a Young diagram with a single row of $k$ boxes. 
The resulting multiplicity is given by the coefficient of $q^k$ in
\begin{eqnarray}
Z_{SH} (q;\Lambda_1)&=&( 1- q ) ~  q^{\sum_{i} c_i(c_i-1)\over 2} ~ \prod_b{1\over (1-q^{h_b})}\cr
&=&\sum_{k}q^k Z_{SH}^k (\Lambda_1)\, . \label{Mult}
\end{eqnarray}
The subscript SH denotes ``symmetrized hook'' and it refers to the fact that we have taken the symmetrized 
($\Lambda_2=[k]$) tensor product of $k$ copies of the hook representation $V_H$.
Here $c_i$ is the length of the $i$'th column in $ \Lambda_1$, $b$ runs over boxes in the Young diagram $\Lambda_1$ 
and $h_b$ is the hook length of the box $b$. 
Evaluating this formula for the antisymmetric representations, for which $\Lambda_1$ is a single column, gives\cite{BHR2}
\bea
{q^{n(n-1)}\over (1-q^2)\cdots (1-q^n)}\, .\label{rres}
\eea
After accounting for the dimension of $n$ elementary fermion fields and reinstating $x$ and $y$, (\ref{rres}) is in complete 
agreement with (\ref{LTCountt}) confirming that the number of polynomials in the complex variables $z^I$ matches
the number of leading twist primary operators.

Now that we have verified that the number of translation invariant, holomorphic polynomials in the antisymmetric
representation of  $S_n$ agrees with the counting of leading twist primaries, we can move on to construction
formulas for these primaries.
Indeed, the relevant polynomials are given by acting with a projector onto the antisymmetric representation,
on the hook variables. This polynomial multiplies an anticommuting tensor product of Grassman valued 
constant spinors.
The projector from the tensor product of $k$ copies of the hook onto the antisymmetric representation of $S_n$ is
\begin{eqnarray}
   P_{(1^n)}={1\over n!}\sum_{\sigma\in S_n}{\rm sgn}(\sigma )\Gamma_k (\sigma)\,,\label{IdProj}
\end{eqnarray}
where ${\rm sgn}(\sigma )$ is the signature of permutation $\sigma$.
When acting on a product of variables, say $Z^{a_1}Z^{a_2}\cdots Z^{a_k}$ we have
\begin{eqnarray}
  \Gamma_k (\sigma)=\Gamma_{H}(\sigma)\otimes\cdots\otimes\Gamma_{H}(\sigma)\,,
\end{eqnarray}
where on the right hand side we take a tensor product (the usual Kronecker product) of $k$ copies of the matrices of 
the hook representation of $S_n$.
Our construction formula is
\bea
{1\over n!}\sum_{\sigma\in S_n}{\rm sgn}(\sigma )\Gamma_k (\sigma)_{a_1 a_2\cdots a_k,b_1 b_2\cdots b_k}
Z^{b_1}Z^{b_2}\cdots Z^{b_k}(\zeta_1\otimes\zeta_2\cdots\otimes\zeta_n)_{\alpha_1\cdots\alpha_n}\, .
\label{LTconstruct}
\eea
The above formula produces an expression of the form $\sum_i \hat{n}_i P_i(Z)$ where $\hat{n}_i$ are 
unit vectors inside the carrier space of $V_H^{\otimes k}$ and $P_i(Z)$ are the polynomials that correspond to
primary operators.
To translate polynomials into momenta, the formula \cite{deMelloKoch:2017dgi}
\bea
  z^k\quad\leftrightarrow\quad {(-1)^k P^k\over 2^k k!}\,,
\eea
that we derived above, is very useful.
We will now give some examples of polynomials obtained from formula (\ref{LTconstruct}).
We will also translate these polynomials into primary operators.

If we consider $n=2$ fields, there is a single hook variable given by $Z=z_1-z_2$.
To find a polynomial that is antisymmetric under swapping $1\leftrightarrow 2$, we must raise $Z$ to an
odd power.
Thus, we find that primaries for the fermion fields correspond to the polynomials
\bea
(z_1-z_2)^{2s+1}=\sum_{k=0}^{2s+1}{(2s+1)!\over k!(2s-k+1)!}(-1)^k z_1^{2s-k+1}z_2^k\, .
\eea
Translating the polynomial variables into momenta we find the following primary
\bea
   |\psi\rangle =\sum_{k=0}^{2s+1} {(-1)^k\over ((2s-k+1)! k!)^2}
     \, P^k |{3\over 2},{1\over 2},0\rangle\otimes\, P^{2s-k+1}|{3\over 2},{1\over 2},0\rangle\,,   \label{n2Primaries}
\eea
where, because our fields are fermions, we have
\bea
|{3\over 2},{1\over 2},0\rangle_1\otimes\, |{3\over 2},{1\over 2},0\rangle_2=-
|{3\over 2},{1\over 2},0\rangle_2\otimes\, |{3\over 2},{1\over 2},0\rangle_1\, .
\eea
Thus, our expression for the fermionic primaries built from two fields are
\bea
\sum_{k=0}^{2s+1} {(-1)^k\over ((2s-k+1)! k!)^2}
     \, (\partial_1+i\partial_2)^k \psi (x) (\partial_1+i\partial_2)^{2s-k+1}\psi (x)\,,
\eea
which exactly matches the form of the higher spin currents\cite{Craigie:1983fb,Giombi:2017rhm}.

For $n=3$ fields it is easy to see that
\bea
(z_1-z_2)(z_1-z_3)(z_2-z_3)\,,
\eea
is holomorphic, translation invariant and in the antisymmetric representation of $S_3$.
The corresponding primary operator can be simplified to
\bea
\psi (x) (\partial_1+i\partial_2) \psi (x) (\partial_1+i\partial_2)^{2}\psi (x)\, .
\eea
It is not difficult to see that this operator is indeed annihilated by $K_\mu$, as discussed in
Appendix \ref{Prmrs}.

\subsection{Extremal Primaries}

In this section we will consider the construction of extremal primaries, which correspond to polynomials in two 
holomorphic coordinates, $z$ and $w$.
The identification of these polynomials with the extremal primaries is again established by showing agreement of
the counting of these polynomials with the counting of extremal primaries.
We will characterize these polynomials by two degrees, one for $Z$ and one for $W$.
Polynomials of degree $k$ in $Z$ and of degree $l$ in $W$ belong to a subspace of 
$V_H^{\otimes k}\otimes V_H^{\otimes l}$ of $S_n$.
The relevant decompositions in terms of $S_n\times S_k$ irreducible representations are
\begin{eqnarray}\label{decompsSnSkSl} 
&& V_H^{ \otimes k } = \bigoplus_{ \Lambda_1 \vdash n , \Lambda_2 \vdash k } V_{ \Lambda_1}^{  (S_n)} \otimes V_{ \Lambda_2}^{ (S_{k} ) } \otimes V^{ Com ( S_n \times S_k )}_{ \Lambda_1 , \Lambda_2} \cr 
&& V_H^{ \otimes l  } = \bigoplus_{ \Lambda_3 \vdash n , \Lambda_4 \vdash l  }  V_{ \Lambda_3}^{  (S_n)} \otimes V_{ \Lambda_4}^{ (S_{l} ) } \otimes V^{ Com ( S_n \times S_l )}_{ \Lambda_3 , \Lambda_4} \, .
\end{eqnarray}
The tensor product $V_H^{\otimes k} \otimes V_{H}^{\otimes l}$ is a representation of 
\bea\label{theAlgebra}  
{\mathbb C} ( S_n ) \otimes {\mathbb C} ( S_k ) \otimes {\mathbb C} ( S_n ) \otimes {\mathbb C} ( S_l ) \, .
\eea
The $Z$ and $W$ variables are commuting so that $\Lambda_2\otimes\Lambda_4=[k]\otimes [l]$ is the trivial representation 
of $S_k\times S_l$.
The multiplicity with which a given $S_n\times S_k$ irrep $(\Lambda_1,\Lambda_2)$ appears is given by the
dimension of the irreducible representation of the commutants $Com ( S_n \times S_l )$ in $V_H^{\otimes k}$.
Recall that since our polynomials multiply a product of anticommuting Grassman spinors, we want to project to states in 
$V_{H}^{\otimes k}\otimes V_H^{\otimes l}$ which are in the totally antisymmetric irreducible representation of the 
diagonal ${\mathbb C} ( S_n ) $ in the algebra (\ref{theAlgebra}). 
This constrains $\Lambda_3=\Lambda_1^T$. 
Thus we find that the number of $ S_k \times S_l$ invariants and $S_n$ antisymmetric representations is 
\begin{eqnarray}
\sum_{\Lambda_1\vdash n} {\rm Mult}(\Lambda_1^T,[k];S_n\times S_k)\,\, {\rm Mult}(\Lambda_1,[l];S_n\times S_l)\, . 
\end{eqnarray}
Thus, for the number of primaries constructed using the variables $z_i,w_i$ we get 
\begin{eqnarray} 
\sum_{ \Lambda_1 \vdash n } Z_{SH}^k  ( \Lambda_1 )
Z_{SH}^l ( \Lambda_1^T )\, .
\end{eqnarray}
The above integer gives the number of primaries in the free fermion CFT, of weight ${3n\over 2}+k+l$, with spin
$(J_1^3,J_2^3)=({k+l+n\over 2},{k-l\over 2})$.
The generating function $ Z^{\rm ext}_n(s,x,y)$ which encodes all $k,l$ is given by
\bea\label{genZzw} 
   Z^{\rm ext}_n(s,x,y)=s^{3n\over 2}x^{n\over 2}
\sum_{\Lambda\vdash n}Z_{SH}(s\sqrt{xy},\Lambda)Z_{SH}(s\sqrt{x\over y},\Lambda^T)\,,
\eea
where $\Lambda$ is a partition of $n$ and we can use the formula (\ref{Mult}). 
It is straight forwards to check, for example, that
\bea
Z^{\rm ext}_n(s,x,y)&=&s^{9\over 2}x^{3\over 2}
\Big(
Z_{SH}(s\sqrt{xy},{\tiny\yng(3)})Z_{SH}(s\sqrt{x\over y},{\tiny\yng(1,1,1)})
+Z_{SH}(s\sqrt{xy},{\tiny\yng(2,1)})Z_{SH}(s\sqrt{x\over y},{\tiny\yng(2,1)})\cr
&+&Z_{SH}(s\sqrt{xy},{\tiny\yng(1,1,1)})Z_{SH}(s\sqrt{x\over y},{\tiny\yng(3)})\Big)
\eea
reproduces (\ref{Z3}).

For $n=3$ fields, it is easy to see that the polynomials
\bea
w_3 (z_2-z_1) + w_2 (z_1 - z_3) + w_1 (z_3 - z_2)
\eea
and
\bea
&&2 w_1 w_2 z_1^2 - w_2^2 z_1^2 - 2 w_1 w_3 z_1^2 + w_3^2 z_1^2 - 2 w_1^2 z_1 z_2 + 
 2 w_2^2 z_1 z_2 + 4 w_1 w_3 z_1 z_2 - 4 w_2 w_3 z_1 z_2\cr\cr
&& + w_1^2 z_2^2 - 2 w_1 w_2 z_2^2 + 2 w_2 w_3 z_2^2 - w_3^2 z_2^2 + 2 w_1^2 z_1 z_3 - 
 4 w_1 w_2 z_1 z_3 + 4 w_2 w_3 z_1 z_3 - 2 w_3^2 z_1 z_3\cr\cr
&&+ 4 w_1 w_2 z_2 z_3- 2 w_2^2 z_2 z_3 - 4 w_1 w_3 z_2 z_3 + 2 w_3^2 z_2 z_3 - w_1^2 z_3^2 + 
 w_2^2 z_3^2 + 2 w_1 w_3 z_3^2 - 2 w_2 w_3 z_3^2\cr\cr
&&
\eea
are holomorphic, translation invariant and in the antisymmetric representation of $S_3$.
To translate these polynomials into primary operators, we use the dictionary
\bea
  z^k w^l \quad\leftrightarrow\quad {(-1)^{k+l} P_z^k P_w^l\over 2^{k+l} (k+l)!}\, .\qquad  \label{Dictionary}
\eea
After a little work we finally obtain the following two primary operators
\bea
\psi_1  = \psi (0) P_z\psi (0) P_w\psi(0)  \label{Primary1}
\eea
and
\bea
 \psi_2  = \frac{1}{3}P_w P_z^2 \psi (0) P_w\psi (0) \psi(0)+ \frac{1}{3}P_z \psi (0) P_w^2 P_z\psi (0) \psi(0)\cr\cr
+{1\over 4}P_w^2 \psi (0) P^2_z\psi (0) \psi(0)+ 2 P_w P_z \psi (0) P_z\psi (0) P_w\psi(0)\, . \label{Primary2}
\eea
In the appendix we verify that these operators are annihilated by the special conformal transformations.  

\section{Geometry}

In this section we comment on the permutation orbifolds relevant for the combinatorics of the fermion primaries.
The leading twist primaries are holomorphic polynomials in $n$ complex variables.
We mod out by translations and restrict to the antisymmetric representation of $S_n$, so that the leading twist
primaries correspond to holomorphic polynomial functions on
\bea
({\mathbb C})^n/({\mathbb C}\times S_n)\, .
\eea
A very similar argument shows that extremal primaries correspond to holomorphic polynomial functions on
\bea
({\mathbb C})^{2n}/({\mathbb C}^2\times S_n)\, .
\eea
We will now argue that the Hilbert series of the fermionic primaries are counted by palindromic Hilbert series,
suggesting that they are Calabi-Yau. We leave a more detailed study of these issues for the future. 
A palindromic Hilbert series obeys
\bea
Z^{\rm ext}_n ( q_1^{-1} , q_2^{-1}  )  = ( q_1 q_2)^{ n-1 }Z^{\rm ext}_n ( q_1 , q_2 )\, .
\eea
Our Hilbert series $Z^{\rm ext}_n (q_1,q_2)$ enjoy this transformation property.
To demonstrate this, our starting point is the formula
\bea 
 Z^{\rm ext}_n ( q_1 , q_2 ) = s^{3n\over 2}x^{n\over 2} 
\sum_{ \Lambda \vdash n } Z_{SH}  ( q_1  , \Lambda ) Z_{ SH} ( q_2 , \Lambda^T )\,,
\eea
where we have introduced the variables $ q_1 = s \sqrt{ xy } , q_2 = s \sqrt{ x /y} $. 
This has the property $ Z_{n}^{\rm ext} ( q_1 , q_2 ) = Z_n^{\rm ext} ( q_2 , q_1 )$. 
This follows because exchange of $q_1 , q_2 $ amounts to the inversion of $y$, and by using the 
identity \cite{deMelloKoch:2017dgi}
\bea
Z_{SH} ( q^{-1} , \Lambda )  = (-q)^{ n-1} Z_{SH} ( q , \Lambda^T )\, . \label{PalZ}
\eea
Using this result we find
\bea 
Z^{\rm ext}_n ( q_1^{-1} , q_2^{-1}  ) & = &  s^n
( q_1 q_2)^{ n-1 }  \sum_{ \Lambda \vdash n } Z_{SH}  ( q_1  , \Lambda^T  ) Z_{ SH} ( q_2 , \Lambda ) \cr 
& = & s^n ( q_1 q_2)^{ n-1 }  \sum_{ \Lambda \vdash n } Z_{SH}  ( q_1  , \Lambda ) Z_{ SH} ( q_2 , \Lambda^T )\cr
& = &  ( q_1 q_2)^{ n-1 }Z^{z,w}_n (q_1,q_2)\, .\label{palin}
\eea
The results of section (4.3) of \cite{deMelloKoch:2017dgi} now imply that the Hilbert series 
$G_n^{\rm ext}(s,x,y)$ also exhibit the palindromy property.

\section{ Summary and Outlook } 

Previous studies \cite{deMelloKoch:2017dgi} have explained how to  map the algebraic problem of constructing primary 
fields in  the quantum field theory of a free scalar field $\phi$ in four dimensions to one of finding polynomial functions 
on $({\mathbb R}^4 )^n$ that are harmonic, translation invariant and which are in the trivial representation of $S_n$.
In this article, we have extended this construction to describe primary fields in the free quantum field theory of a
single Weyl fermion.
Concrete results achieved with this new point of view include a counting formula for the complete
set primary fields, explicit counting formulas (Hilbert series) for counting special classes of primaries, as well as
detailed construction formulas for these primary operators. 
We have also established the palindromy of the Hilbert series

One weak point in our analysis, that warrants further study, is the treatment of the constraint coming from the equation
of motion.
We have simply demonstrated that polynomials holomorphic in the complex variable $z$ and $w$, times the spinor $\zeta$
with maximal $J_1^3$ eigenvalue, solve the equation of motion constraint.
Our results have been further verified by checking that the numbers of polynomials constructed from a singe complex variable
match the numbers of leading twist primaries, that the number of polynomials constructed from two complex variables
match the number of extremal primaries and further that when
the polynomials are translated back into the operator language, that we do indeed obtain operators annihilated by $K_\mu$.
It would however be nice to perform a detailed analysis of the equation of motion constraint, which has to be carried out
before the complete class of primaries can be constructed.

There is an immediate generalization of our study which should be tackled.
CPT invariance implies we need both right handed and left handed fermions.
As a concrete example, the Hilbert space for a single Weyl fermion is
\bea
{\cal H}=\bigoplus_{n=0}^\infty {\rm Asym}^n (W_+\oplus W_-)
\eea
Can the methods developed in this article be used to study the above Hilbert space?
In this situation $M=s^D x^{J_1^3}y^{J_2^3}=M_+\oplus M_-$, with $M_+$ associated to $W_+$ and $M_-$
associated to $W_-$.
The basic identity we are using becomes
\bea
  \det (1+ t_+M_+ \oplus t_- M_-)&=&\sum_{n_+,n_-=0}^\infty t_+^{n_+} t_-^{n_-}\chi_{n_+,n_-}(M)\cr
&=&\prod_{q_+,q_-=0}^\infty\prod_{a=-{q+1\over 2}}^{q+1\over 2}\prod_{b=-{q\over 2}}^{q\over 2}
(1+t_+s^{{3\over 2}+q}x^a y^b)(1+t_-s^{{3\over 2}+q}x^b y^a)\cr
&&
\eea
where $\chi_{n_+,n_-}(M)$ is the antisymmetrized product of $n_+$ copies of $M_+$ and $n_-$ copies of $M_-$.
By expressing $\chi_{n_+,n_-}(M)$ as a sum of characters of irreducible representations of the conformal group, we
learn what primaries can be constructed from a product of $n_+$ left handed and $n_-$ right handed Weyl fermions.
To achieve this decomposition, the methods of section 2 can be employed.
To obtain simple and explicit results, one can again consider restricting the resulting counting formulas to special
classes of primaries. 
For a given pair of integers $(n_+,n_-)$, we can define both the leading twist (we maximize both $J_1^3$ and $J_2^3$ 
at a given dimension) and the extremal (we maximize $J_1^3$ or $J_2^3$ at a given dimension) classes of primaries.
For these classes, it would be very interesting to see if a symmetric group interpretation of the counting can be developed,
along the lines of sections 3.2 and 3.3. 
We are currently exploring this promising possibility and hope to return to it in the near future.
A symmetric group interpretation of the counting would immediately suggest detailed construction formulas for the 
associated primaries.

Given that the counting for a Weyl fermion and a scalar field have been carried out, it is natural to ask if one
can assemble this counting to give the counting of superconformal primaries.
The simplest starting point would be a free boson plus fermion theory, where the counting of this paper for the fermion 
and of \cite{deMelloKoch:2017dgi} for the boson, would be directly applicable.
Other generalizations of the current work would include studies of CFTs which include gauge fields. 
Note that early constructions of primary fields in the $SL(2)$ sector (leading twist primaries)  were 
performed in the context of  deep inelastic scattering in QCD (see for example \cite{BKM03}), suggesting
that the free limit of QCD maybe a good starting point. 
Another natural question is the explicit enumeration and construction of superconformal primary fields in ${\cal N}=4$ 
SYM, which will give a better understanding of the dual $ AdS_5 \times S^5$ background. 
Finally, our results maybe useful when considering correlators involving the extremal primary fields at the fixed point of 
the Gross-Neveu model in $2+\epsilon$ dimensions.
In particular, one could attempt to determinae the anomalous dimensions of these fields , using the
techniques of \cite{Rychkov:2015naa,Basu:2015gpa,Ghosh:2015opa,Raju:2015fza,Nii:2016lpa}. 

{\vskip 0.5cm}
\noindent
\begin{centerline} 
{\bf Acknowledgements}
\end{centerline} 

This work is supported by the South African Research Chairs
Initiative of the Department of Science and Technology and National Research Foundation
as well as funds received from the National Institute for Theoretical Physics (NITheP).
HJR is supported by a Claude Leon Foundation postdoctoral fellowship.
We are grateful for useful discussions to Shinji Hirano and Sanjaye Ramgoolam.

\appendix

\section{Primaries examples}\label{Prmrs}

In this Appendix we will collect a few details on the translation from polynomials to primary operators and then
test, for a few examples, that the primaries obtained are indeed annihilated by $K_\mu$.

\subsection{Dictionary}

We first show that the appropriate way to translate between polynomials and operators is given by (\ref{Dictionary}).  
We again make use of the (Euclidean) representation
\begin{equation}
P_\mu = x^2 \partial_\mu - 2 x_\mu x\cdot \partial +3 x_\mu -2 x_\nu {\cal M_{\mu\nu}} \, .
\end{equation}
We consider a polynomial in $P_z =P_2+iP_1= \epsilon_z \cdot P$ and $P_w =P_3-iP_4= \epsilon_w \cdot P$ 
acting on the constant spinor $\zeta$. 
The $\epsilon$'s obey the following identities
\begin{eqnarray}
\epsilon_z \cdot \epsilon_z & = & 0 = \epsilon_w \cdot \epsilon_w = \epsilon_z \cdot \epsilon_w  \nonumber \\
\epsilon_z \cdot x = x_2+ix_1 = z \,, && \epsilon_w \cdot x = x_3+ix_4 = w\, .  
\end{eqnarray}
The analysis of this Appendix is for the leading twist and extremal primaries.
In this case we have fixed the left spin to a maximal value, corresponding to choosing the spinor $\zeta$ with spin up.
Useful formulas to bear in mind are
\begin{eqnarray}
(\epsilon_z)_{\mu}(3 x_\mu -2 x_\nu {\cal M_{\mu\nu}})\zeta &=& 2 z \zeta\,,\cr
(\epsilon_w)_{\mu}(3 x_\mu -2 x_\nu {\cal M_{\mu\nu}})\zeta &=& 2 w \zeta\, .
\end{eqnarray}
Finally, we will also make use of the fact that
\begin{equation}
P_w^k P_z^l\psi (0) \leftrightarrow (-2)^{k+l} (k+l)! w^k z^l\zeta \, .
\end{equation}

\subsection{$n=2$ Example}

Consider the operators given by (\ref{n2Primaries}).  Introduce $K_z = K_2 - i K_1$. 
It is straightforward to verify that $[K_{\bar{z}}, P_z] = 0 = [P_{\bar{z}}, K_z]$ and
\begin{eqnarray}
\left[D, P_z\right] & = & P_z\,, \nonumber \\
\left[D, K_z\right] & = & -K_z\,, \nonumber \\
\left[K_z, P_z\right] & = & - 4D + 4iM_{21} \label{CommRel} \\
\left[M_{21}, K_z\right] & = & -i K_z \nonumber \\
\left[M_{21}, P_z\right] & = & i P_z\, .  
\end{eqnarray}
We will now argue that $K_{\bar{z}}$ annihilates (\ref{n2Primaries}).  
Using the above algebra we easily find
\begin{equation}
K_z P_z^{m} |\frac{3}{2}, \frac{1}{2}, 0 \rangle = -4 P_z^{m-1}(m^2 - m (1 - D + i M_{21})) |\frac{3}{2}, \frac{1}{2}, 0 \rangle 
\end{equation}
Consequently the action of $K_z$ on the state (\ref{n2Primaries}) yields
\begin{eqnarray}
K_z|\psi\rangle & = & -4\sum_{k=1}^{2s+1} {(-1)^k k^2 \over ((2s-k+1)! k!)^2}
     \, P^{k-1} |{3\over 2},{1\over 2},0\rangle\otimes\, P^{2s-k+1}|{3\over 2},{1\over 2},0\rangle \nonumber \\
			 & & - 4\sum_{k=0}^{2s} {(-1)^k(2s - k+1)^2\over ((2s-k+1)! k!)^2}
     \, P^k |{3\over 2},{1\over 2},0\rangle\otimes\, P^{2s-k}|{3\over 2},{1\over 2},0\rangle \nonumber \\
		& = & 0
\end{eqnarray}
Next, it is straight forward to verify that
\begin{eqnarray}
K_3 P_z^{m}|0\rangle & = & im P_z^{m-1}(M_{31} - i M_{32})|0\rangle \nonumber \\
K_4 P_z^{m}|0\rangle & = & im P_z^{m-1}(M_{41} - i M_{42})|0\rangle.  
\end{eqnarray}
The operators $M_{31} - i M_{32}$ and $M_{41} - i M_{42}$ are raising operators for the right spin.
Since the state $|\frac{3}{2}, \frac{1}{2}, 0 \rangle$ has vanishing right spin, we have
\begin{eqnarray}
i M_{21} \ |\frac{3}{2}, \frac{1}{2}, 0 \rangle & = & 
i M_{34} \ |\frac{3}{2}, \frac{1}{2}, 0 \rangle = {1\over 2} \ |\frac{3}{2}, \frac{1}{2}, 0 \rangle \nonumber \\
(M_{31} - i M_{32}) \ |\frac{3}{2}, \frac{1}{2}, 0 \rangle  & = & 0 =  (M_{41} - i M_{42}) \ |\frac{3}{2}, \frac{1}{2}, 0 \rangle.  \label{RefStateRep}
\end{eqnarray}
It now follows that $K_3$ and $K_4$ annihilate (\ref{n2Primaries}), completing the demonstration that (\ref{n2Primaries}) is
indeed a primary operator.  

\subsection{$n=3$ Examples}

We will show that the operators (\ref{Primary1}) and (\ref{Primary2}) are annihilated by the special conformal generators.
Define $K_w = K_3 + i K_4$.  
It is straightforwards to evaluate
\begin{eqnarray}
\left[K_w, P_z\right] &=& 2\left(M_{32}+iM_{31}-(M_{41}-iM_{42})\right) \equiv 4iM_{wz} \nonumber \\
\left[M_{wz}, P_w\right] & = & i P_z \nonumber \\
\left[M_{wz}, P_z\right] & = & 0
\end{eqnarray}
and
\begin{eqnarray}
\left[K_z, P_w\right] &=& 2(M_{41}+iM_{42}-(M_{32}-iM_{31})) \equiv 4i M_{zw} \nonumber \\
\left[M_{zw}, P_z\right] & = & i P_w \nonumber \\
\left[M_{zw}, P_w\right] & = & 0
\end{eqnarray}
To interpret these commutators, note that $P_z$ has spin $({1\over 2},{1\over 2})$ and $P_w$ has spin
$({1\over 2},-{1\over 2})$.
Thus, $M_{wz}$ and $M_{zw}$ are raising/lowering operators of the right spin.
Since our fermion field has vanishing right spin it is clear that
\begin{eqnarray}
M_{zw} |\frac{3}{2}, \frac{1}{2}, 0 \rangle=M_{wz} |\frac{3}{2}, \frac{1}{2}, 0 \rangle & = & 0
\end{eqnarray}
which implies the identities
\bea
K_z P_w^n P_z^m |\frac{3}{2}, \frac{1}{2}, 0 \rangle 
= -4(n m + m^2 )P_w^n P_z^{m-1} |\frac{3}{2}, \frac{1}{2}, 0 \rangle\cr
K_w P_w^n P_z^m |\frac{3}{2}, \frac{1}{2}, 0 \rangle 
= -4(n m + n^2 )P_w^{n-1} P_z^{m} |\frac{3}{2}, \frac{1}{2}, 0 \rangle
\eea
It now follows that
\bea
K_z \psi_1 = -4 \psi(0) \psi(0) P_w \psi(0) = 0\cr
K_w \psi_1 = -4 \psi(0) P_z \psi(0) \psi(0) = 0
\eea
where we used the Grassman statistics of the field.  
For the action on $\psi_2$ we find 
\begin{eqnarray}
-{1\over 4}K_z \psi_2 & = & \frac{1}{3} (6) P_w P_z \psi(0) P_w \psi(0) \psi(0) 
+ \frac{1}{3} (3) P_z \psi(0) P_w^2 \psi(0) \psi(0) \nonumber \\
& & + \frac{1}{4}(4) P_w^2 \psi(0) P_z \psi(0) \psi(0) + 2 P_w P_z \psi(0) \psi(0) P_w \psi(0) 
 = 0\cr
-{1\over 4}K_w \psi_2 & = & \frac{1}{3} (3) P_z^2 \psi(0) P_w \psi(0) \psi(0) 
+ \frac{1}{3} (6) P_z \psi(0) P_w P_z \psi(0) \psi(0) \nonumber \\
& & + \frac{1}{4}(4) P_w \psi(0) P_z^2 \psi(0) \psi(0) + 2 P_w P_z \psi(0) P_z \psi(0) \psi(0) 
 = 0
\end{eqnarray}
again after using the Grassman nature of the field.  
This completes the demonstration that $\psi_1$ and $\psi_2$ are primary operators.

{} 

\end{document}